\documentclass[reprint,superscriptaddress,amsmath,amssymb,aps,showpacs,nofootinbib]{revtex4-1}

\usepackage[utf8]{inputenc}
\usepackage{graphicx}
\usepackage{amsmath}
\usepackage{amssymb}
\usepackage[english]{babel}
\usepackage[exponent-product=\cdot,scientific-notation=false,separate-uncertainty=false,per-mode=symbol]{siunitx}
\usepackage[textsize=small]{todonotes}
\usepackage{booktabs}
\usepackage{color}
\usepackage{xcolor}
\usepackage{ulem}

\begin{document}

\title{The anisotropy in the optical constants of quartz crystals for soft X-rays}
\author{A. Andrle}
\email{anna.andrle@ptb.de}
\affiliation{Physikalisch-Technische Bundesanstalt (PTB), 
Abbestr. 2-12, 10587 Berlin, Germany}

\author{P. H\"{o}nicke}
\affiliation{Physikalisch-Technische Bundesanstalt (PTB), 
Abbestr. 2-12, 10587 Berlin, Germany}

\author{J. Vinson}
\affiliation{National Institute of Standards and Technology (NIST), Gaithersburg, Maryland, USA}

\author{R. Quintanilha}
\affiliation{ASML Netherlands B.V. (ASML), Netherlands }

\author{Q. Saadeh}
\affiliation{Physikalisch-Technische Bundesanstalt (PTB), 
Abbestr. 2-12, 10587 Berlin, Germany}

\author{S. Heidenreich}
\affiliation{Physikalisch-Technische Bundesanstalt (PTB), 
Abbestr. 2-12, 10587 Berlin, Germany}

\author{F. Scholze}
\affiliation{Physikalisch-Technische Bundesanstalt (PTB), 
Abbestr. 2-12, 10587 Berlin, Germany}

\author{V. Soltwisch}
\affiliation{Physikalisch-Technische Bundesanstalt (PTB), 
Abbestr. 2-12, 10587 Berlin, Germany}

\begin{abstract}
The refractive index of a y-cut SiO$_2$ crystal surface is reconstructed from polarization dependent soft X-ray reflectometry measurements in the energy range from 45 eV to 620 eV. Due to the anisotropy of the crystal structure in the (100) and (001) directions, we observe a significant deviation of the measured reflectance at the Si-L$_{2,3}$ and O-K absorption edges. The anisotropy in the optical constants reconstructed from these data is also confirmed by ab initio Bethe-Salpeter Equation calculations of the O-K edge. This new experimental dataset expands the existing literature data for quartz optical constants significantly, particularly in the near-edge regions.
\end{abstract}

\maketitle
  
\section{Introduction}
Silicon dioxide (SiO$_2$) is very well known for its polymorphism~\cite{BRUCKNER1970}. One of its crystalline forms is called quartz and is classified into different types (I - IV) depending on the manufacturing process and the resulting impurities~\cite{BRUCKNER1970, Kitamura2007}. Quartz glass is used today in a wide variety of applications, from simple laboratory glassware and optics to semiconductor manufacturing and lithography photomasks. In the field of optoelectronics and Micro-Electro-Mechanical Systems (MEMS), quartz, similar to pure silicon due to its physical properties, is often used as a carrier material for various mirrors, nanostructures or other functional surface features.\\
For the development of new optical components with tailor-made properties - {\it e.g.} maximum reflectivity in a certain wavelength range for a mirror - reliable material knowledge is required. The determination of accurate optical constants is a key factor for the modeling of light-matter interactions and for the design of novel optical devices. In the literature many different measurements of the optical constants of amorphous SiO$_2$ exist~\cite{yanagihara_soft_1988,Tripathi2002,Filatova_1996}. A good overview of measurements for amorphous silica glass from EUV to infrared can be found in Kitamura et al.~\cite{Kitamura2007}.\\

The optical constant or dielectric permittivity, however, is not always only dependent on the wavelength, but for certain materials also on the wave vector. This dependence is also called spatial dispersion or optical anisotropy and can also occur, for example, in a perfect cubic crystal that should be completely isotropic. The effect of optical anisotropies near the absorption edges was theoretically postulated by Ginzburg~\cite{Ginzburg_1958} already in 1958 and is often referred to as birefringence. In the vicinity of a strong absorption edge, the dielectric function and thus the refractive index increases strongly. The wavelength within the medium can thus shrink to the order of the lattice constant.These anisotropies have been observed in various materials~\cite{Yu_1971,Pastrnak_1971,Letz_2003} in the past and have also been confirmed for quartz at the O-K edge by X-ray absorption near edge spectroscopy~\cite{taillefumier_x-ray_2002}.\\

The primary interactions in the low-energy range of X-rays are photoabsorption and coherent scattering. These interactions can be well described with the complex atomic scattering factors $f(0)$ (the Fourier transformation of the charge distribution) assuming that the individual atoms scatter independently. The total coherent scattering intensity can then be described as the sum of the scattering amplitudes of the individual atoms. For long wavelengths, compared to atomic dimensions and scattering amplitudes which are in phase, the angular dependence of the scattering factor disappears. The interaction of X-ray and matter is then often described by optical constants such as the complex refractive index $\tilde{n}$ or dielectric constant $\sqrt{\epsilon}$ (for non magnetic materials):
\begin{equation}
    \tilde{n} = 1-\delta - i\beta =  1- \frac{r_0}{2\pi}\lambda^2\sum_{i}n_i f_{i}(0) 
\end{equation}
where $r_0$ is the classical electron radius, $\lambda$ the photon wavelength, $n_i$ the atom number density and $f(0)=f_{1}+if_{2}$ represents the complex atomic form factor. The real part 1-$\delta$ and the imaginary part $\beta$ of the complex refractive index are often referred to as optical constants or the n\&k values of a certain material. This approximation is sufficient for photon energies above 30 eV and far away from the absorption edges~\cite{Henke1993}. To experimentally verify or determine these optical constants, the reflection, the transmission, or the absorption of the material is measured at different wavelengths~\cite{Poelman_2003}. However, transmission experiments in the soft X-ray range usually require very thin and free standing material samples, which are not always available. Therefore, indirect methods (total electron or fluorescence yield) are often used in this spectral range to determine the absorption. The real part of the refractive index can then be calculated using the Kramers-Kronig relationship~\cite{filatova_anisotropy_2002,Soufli1997}. A widely used alternative is the measurement of the specular reflectivity. It is known as X-ray reflectometry (XRR)~\cite{Stoev1997} and is mostly used to determine the thickness of thin films or multilayer systems with sub-nm precision~\cite{Gil2012}.\\

\begin{figure*}[tbp]
\label{fig:theta_E_R}
\centering
\includegraphics[width=0.95\textwidth]{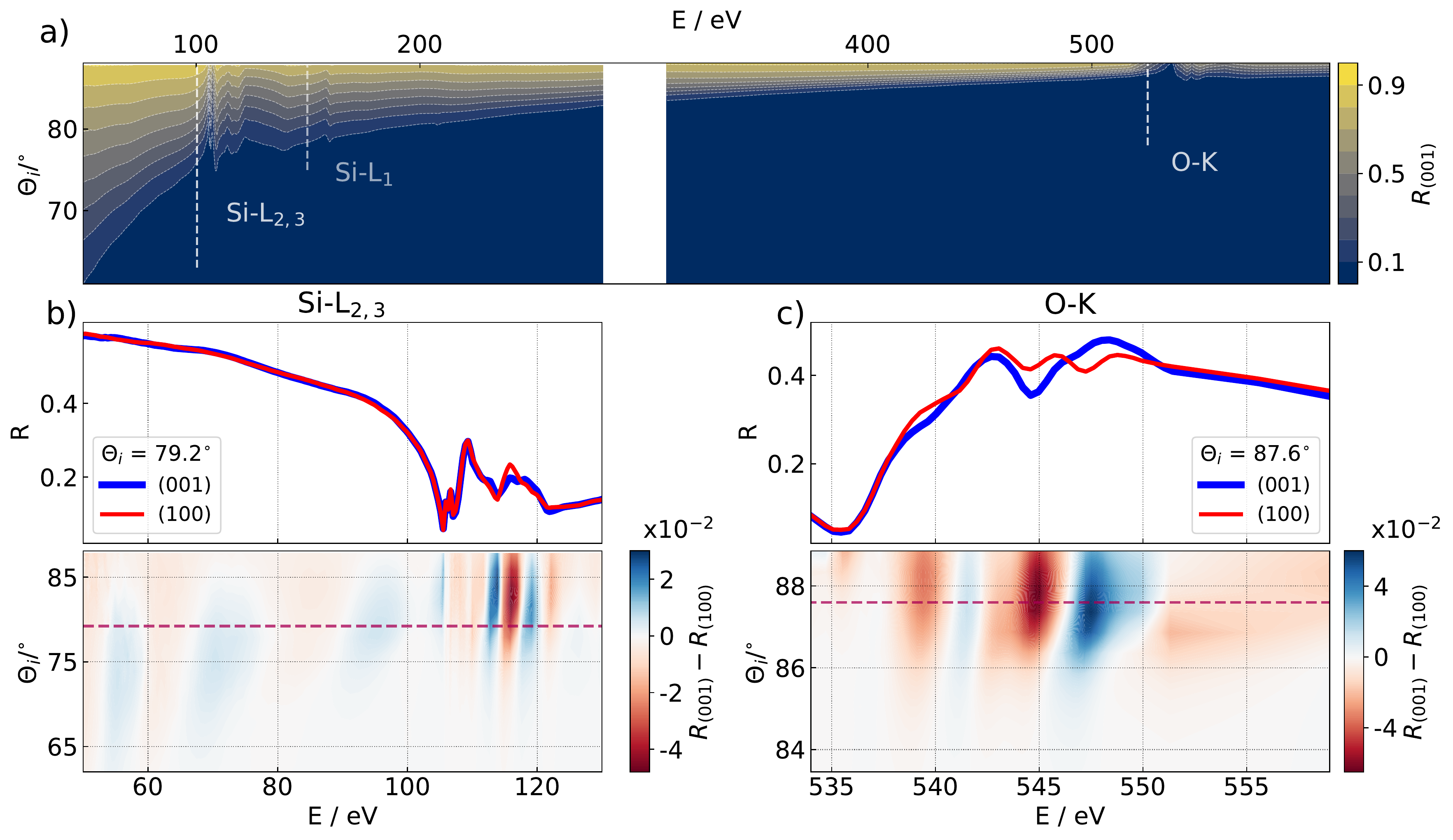}
\caption{a) Specular reflectance map of quartz (y-cut) for the measured $\theta_i$ and photon energy range $E$ in the (001) direction. The white region marks the area not measurable due to the carbon edge. b) and c) Comparison of the measured reflections for the crystal directions (100) and (001) at a fixed angle of incidence $\theta_i$ around the b) Si-L$_{2.3}$ and c) O-K absorption edges. The lower maps in b) and c) show the respective anisotropy maps.}
\end{figure*}

In this paper we evaluated polarization dependent reflection measurements with soft X-rays on a quartz crystal. Our data cover a photon energy range from 45 eV to 620 eV and include two crystal orientations. To limit the numerical effort of the global optimization of optical constants in a large energy range, an adapted meta-heuristic optimization algorithm (Differential Evolution~\cite{Storn1997}) was combined with a quasi-Newton method~\cite{Byrd1995}. The optical constants at the absorption edges reconstructed from the soft X-ray reflectivity measurements confirm the theoretically expected anisotropy of quartz in the complex refractive index. Furthermore, an unexpected anisotropy in front of the Si-L$_{2,3}$ edge can be observed. The observed anisotropy at the O-K edge is also confirmed by ab initio simulations using {\sc ocean} (Obtaining Core Excitations from Ab initio electronic structure and NIST BSE) calculations~\cite{Gilmore_2015,Vinson_2011}.

\section{Experimental Details}
Soft X-ray reflectometry measurements were conducted at the the soft X-ray radiometry beamline, operated by the Physikalisch-Technische Bundesanstalt (PTB), at the electron storage ring BESSY II in Berlin \cite{scholze2001}. The beamline covers the photon energy range from \SI{45}{eV} to \SI{1800}{eV} and is designed to produce a beam with low divergence ($<$\SI{1}{mrad}) with minimal halo. 

The sample was mounted on a 6-axis goniometer in the ellipso-scatterometer under ultrahigh vacuum (UHV) conditions. The angle of incidence $\theta_i$ was aligned with respect to the beam with an uncertainty of 0.01$^{\circ}$. The angle of incidence was varied between 0$^{\circ}$ and 88.7$^{\circ}$ in s-polarization direction from the sample normal. The range of incidence angle was adapted to the different photon energies in order to effectively cover the critical angle range.
The analyzed quartz (type II) surface (y-cut) was additionally aligned with respect to the reference plane ($0^{\circ} \pm 0.08^{\circ}$) of the (001) direction and the (100) direction rotated 90$^{\circ}$ azimuthally to it. The specular reflection was measured with a GaAsP photodiode mounted on a movable detector arm inside the vacuum chamber and normalized to the incoming photon flux.\\
In addition, to resolve the anisotropy of the crystal structure of quartz, the reflection measurements in the near edge region of Si-L$_{2,3}$ and O-K were performed with a energy increment slightly lower then the energy resolution of the beamline\cite{scholze2001} ($\frac{E}{\Delta E} = 1000$). 
The total photon energy range covered for the quartz measurements is between $\SI{45}{eV}$ and $\SI{620}{eV}$.\\
\begin{figure*}[htb]
\label{fig:nk_complete}
\centering
\includegraphics[width=0.95\textwidth]{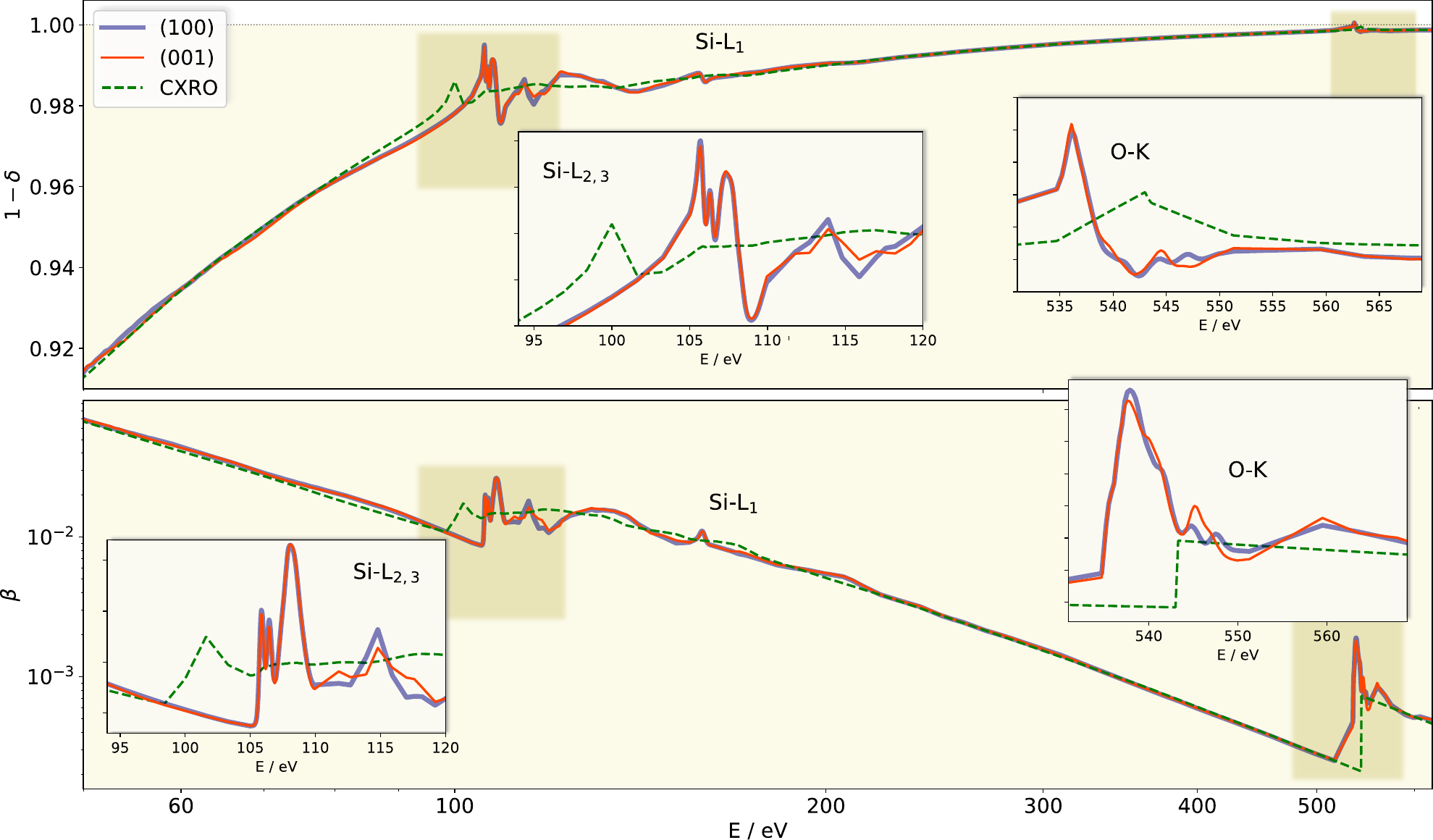}
\caption{Reconstructed refractive index of a quartz crystal. Shown is the $1-\delta$ and $\beta$ part for the ordinary (001) (red line) and extraordinary (100) (blue line) orientation of the crystal compared to the SiO$_2$ CXRO database values (green dashed line)~\cite{Henke1993}. The insets show a magnified view of the reconstructed anisotropy in the absorption edge areas.}
\end{figure*}   

In Figure~\ref{fig:theta_E_R} a) the measured specular intensity as a function of $\theta_i$ and $E$ is shown as a contour map for the (001) direction.  Figure~\ref{fig:theta_E_R} b) and c) show details of the anisotropy at the absorption edges of Si-L$_{2,3}$ and O-K. The line plots at a fixed angle of incidence $\theta_i$ reveal that the measured anisotropy at the absorption edges is not a measurement artifact, since the measurement uncertainty is within the smallest line width. In addition, a weak birefringence below the Si-L$_{2,3}$ absorption edge can also be observed in the anisotropy map in Figure~\ref{fig:theta_E_R} b). 

\section{Reconstruction of the optical constants from reflectivity measurements}
If the complex refractive index $\tilde{n}$ of a material is known, the expected reflectivity from the surface can be calculated as a function of the incidence angle $\theta_i$. The change of the wave vector component $k_z$ at the $j$ interface can be written as:
\begin{equation}
k_{z,j} = \sqrt{(n_j k_0 )^2-\sin^2(\theta_i)k_0^2}    
\end{equation}
with $k_0$ as the incident wave vector. By using the Fresnel coefficients, the reflection and transmission through the medium can then be calculated directly. However, this is not sufficient to describe the measured reflectivities of the quartz crystal. The contamination of the crystal surface with carbon and water must be considered in the simulations, as the sample could not be cleaned in vacuum. The Transfer-Matrix Method provides a method to calculate the specular reflectivity for a multi-layer system depending on the layer thickness and the roughness~\cite{Gibaud2007}. Assuming a Gaussian distribution of roughness and interdiffusion, the modified Fresnel coefficients $\tilde{r}_j$ and $\tilde{t}_j$ for each interface $j$ can then be written~\cite{Croce1976} as:
\begin{align*}
\tilde{r}_{j} &= r_{j}\exp\bigl(-2k_{z,j} k_{z,(j+1)}\sigma_j^{2}\bigr)\: \mathrm{and}\\
\tilde{t}_{j} &= t_{j}\exp\bigl((k_{z,j} - k_{z,(j+1)})^2 \sigma_j^{2}/2\bigr).
\end{align*}
\label{eq:R_k}
The parameter $\sigma_j$ represents the mean square intermixing at the $j$th interface.\\

To stabilize the stratified system and the results for the reconstruction of the optical constants, all measured reflectivities ($\num{2e5}$) at different incidence angles $\theta_i$ and photon energies $E$ were optimized simultaneously. Due to the lack of suitable n\&k models in the EUV spectral range, the challenge in the optimization process arises in the very large number of degrees of freedom. For each $\theta$ scan at at a fixed wavelength an independent refractive index $n$ had to be assumed.\\
In the reconstruction attempts, it quickly became apparent that the use of fast gradient methods for the optimization led to inconsistent results depending on the selected start parameters. This problem of getting stuck in local minima while simultaneously optimizing layer parameters and optical constants is well known~\cite{Cao_94}. More suitable for the global minimization of a problem are heuristic or meta-heuristic methods. The convergence of the differential evolution method (DE)~\cite{Storn1997}, however, is very slow when several hundred degrees of freedom are considered. But the present problem can be easily split up. For a defined geometric model $\vec{p}$ of the layer system there is only one refractive index that can best describe the measured reflectivity. Therefore, the combination of two different optimization methods might be used. In an external optimization, the best combination of geometric parameters are evaluated by means of DE, while a quasi-newton algorithm (L-BFGS-B~\cite{Byrd1995}) determines the corresponding optical constants ($\delta, \beta$), for every layer, inside the objective function $\chi\prime$ of the DE. The objective function can then be written as:
\begin{equation}
\chi\prime (\vec{p}) = \min_{\delta , \beta} \chi(\vec{p},\delta,\beta)\:.   
\end{equation}
The optical constants obtained with this approach, in the energy range from 45 eV to 620 eV, are in good agreement with tabulated values of the databases~\cite{Henke1993,Chantler_2000,Pailik} for SiO$_2$ with an expected density of 2.65 \si{\gram\per\cubic\cm}. However, this perfect agreement is only valid in the regions far away from the absorption edges of silicon (Si-L$_{2,3}$) and oxygen (O-K). The absorption edge positions of the tabulated data are energetically clearly shifted compared to our data, and of course the fine structure is also missing. The insets in Figure 2 highlight the behavior of the refractive index around the absorption edges and allow to visualize the quartz anisotropy.\\

For this result, however, the model of a pure quartz substrate had to be revised. To compensate the effect of the substrate's contamination due to the adsorption of volatile organic materials as the sample was handled at normal laboratory conditions, an ultra-thin carbon like layer with unknown optical and dimensional parameters was considered for the optimization model. It has been reported in the literature that ultra-thin organic coatings can significantly influence the coating thicknesses derived from X-ray reflectometry due to environmental contamination~\cite{Gil2012}. An equivalent problem can also be observed in the reconstruction of the optical constant.\\
The reconstructed surface layer thickness of $\approx 0.6$ nm with an rms roughness of $\approx 0.5$ nm also seems reasonable for the contamination of the quartz surface. This assumption is also confirmed by the following simulations of the O-K edge and shows that modeling errors have a significant influence on the reconstruction of the optical constants in the near edge region.
\newpage

\section{Theoretical modeling of the O-K absorption edge of quartz}
Several groups have already published calculations of X-ray absorption fine structures at the O-K edge~\cite{Gougoussis_2009,taillefumier_x-ray_2002}. In these simulations the core-hole excited state was modeled using density functional theory (DFT). A core-level electron is removed from the absorbing atom, and the electronic system is allowed to relax fully.\\ 
Here we use instead the method of the Bethe-Salpeter equation (BSE). In this equation the ground state is modeled by DFT and the core-hole interactions of the excited state are explicitly calculated. The main difference between the two approaches is that for BSE the electronic relaxation is performed in response to the core-hole by means of a linear response, and for the DFT approach the exchange interaction between electron and hole is approximated.\\

\begin{figure}[tbp]
\centering
\includegraphics[width=0.49\textwidth]{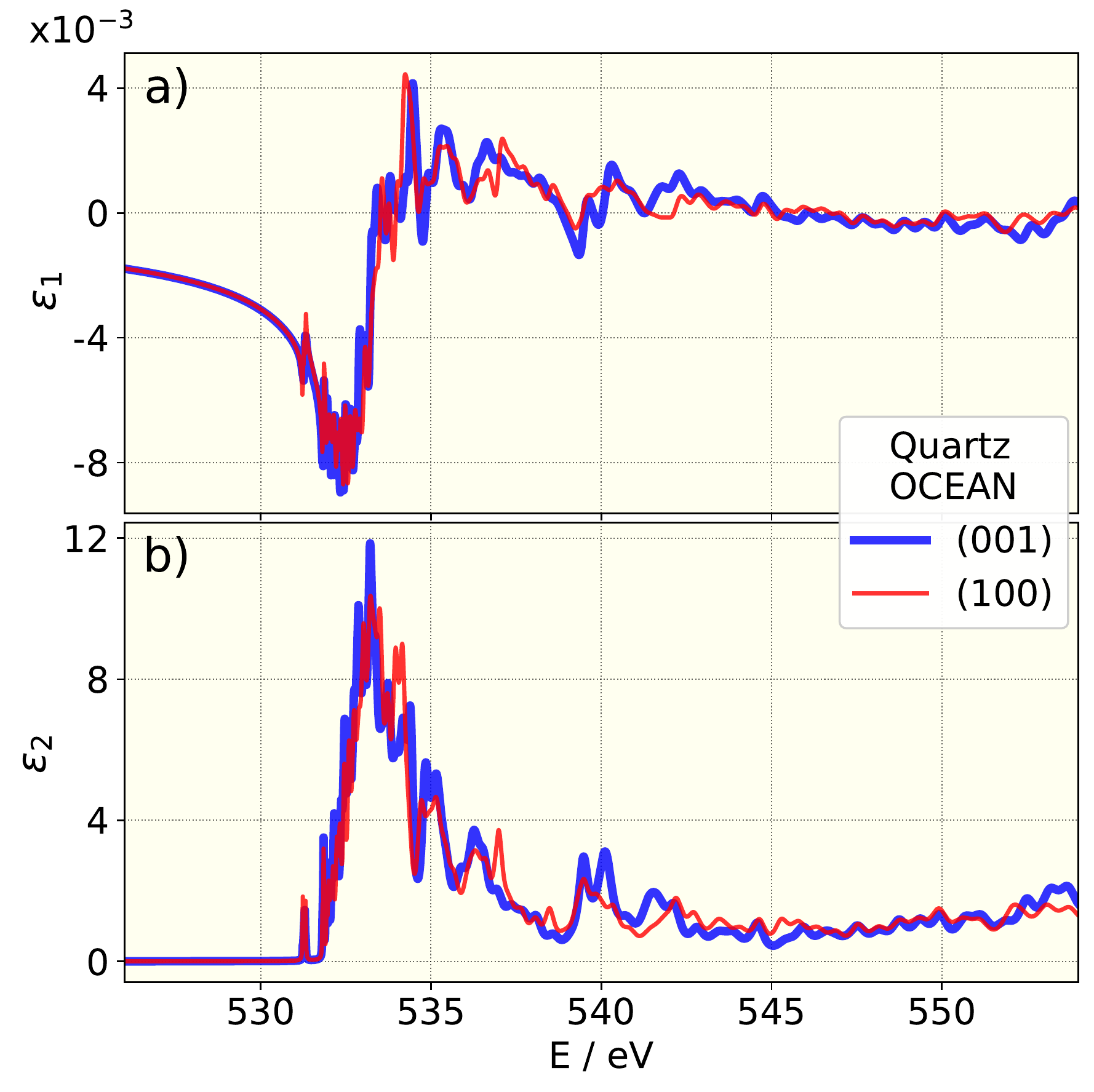}
\caption{Comparison of the {\sc ocean} forward calculation of the permittivity a) $\epsilon_1$ and b) $\epsilon_2$ at the O-K absorption edge for the extraordinary (100) and ordinary (001) orientation of the crystal.}
\label{fig:epsilon_Oedge}
\end{figure}
\newpage
We performed calculations\footnote{We used a plane wave energy cut-off of 55~Ry, and the isotropic dielectric constant $\epsilon_\infty$ was set to 3.8. A $12 \times 12 \times 12$ {\it k}-point grid and 500 bands were used for the BSE final states, while a $4 \times 4 \times 4$ {\it k}-point grid and 150 bands were used for calculating the screening of the core hole.} at the O-K absorption edge using the {\sc ocean} (version 2.5.2)\cite{Gilmore_2015,Vinson_2011} code (see Figure \ref{fig:epsilon_Oedge}).
The electronic ground state was calculated with DFT within the local‐density approximation using the QuantumESPRESSO code\cite{Giannozzi_2017}. For both oxygen and silicon, the pseudopotentials of the Fritz-Haber-Institute (FHI) from the QuantumESPRESSO website were used. The crystal data for quartz was taken from Kihara et al.~\cite{Kihara_1990} for room temperature and cif2cell~\cite{Bj_rkman_2011} was used for conversion.

\begin{figure}[htbp]

\centering
\includegraphics[width=0.49\textwidth]{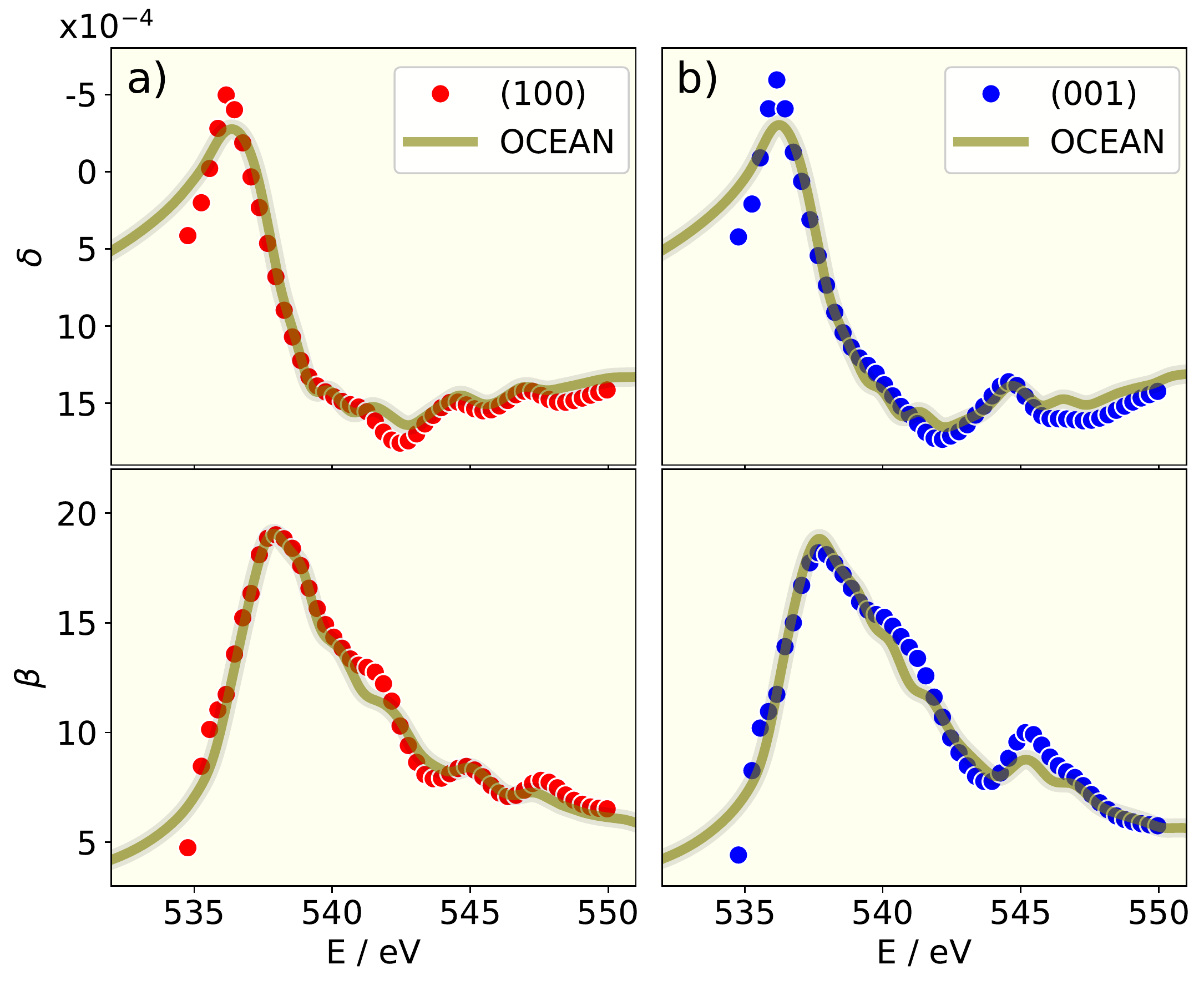}
\caption{Comparison of the reconstructed optical constants ($\delta, \beta$) at the O-K absorption edge for the a) extraordinary (100) and b) ordinary (001) orientation of quartz with the {\sc ocean} simulation of the expected behavior of a quartz crystal. For the reconstruction of the optical constants from the reflectivity measurements a contamination of the surface was modelled.}
\label{fig:nk_Oedge}
\end{figure}
\subsection{\sc{ocean} post-processing }
For a comparison of the calculated complex permittivity $\sqrt{\epsilon} = \sqrt{\epsilon_1 + i\epsilon_2}$ with the experimentally determined optical constants at the oxygen O-K edge, several corrections must be applied to the forward calculation. Due to the limitations of the DFT, a polynomial correction of the energy scale has been applied, which essentially allows a slight modification of the position of each peak. In addition, the lifetime broadening as well as the energy resolution of the beamline must be approximated with a Voigt distribution. The contributions for energetically less bound electrons are considered by Ebel polynomials~\cite{H.Ebel2003}. The corresponding model parameters are determined by means of a least squares optimization. Thereby both $\epsilon_1$ and $\epsilon_2$ are optimized simultaneously with an identical set of parameters. The results of the {\sc ocean} post-processing for the O-K edge are compared against the reconstructed optical constants in Figure~\ref{fig:nk_Oedge}. The shape of $\delta$ and $\beta$ curves in the region of the oxygen edge is already reasonably well reproduced.\\ 
The BSE approach makes a number of approximations which limit its ability to fully reproduce experimental data. First, the atomic nuclei are treated as fixed in their equilibrium position. This neglects both the movement of the atoms in the ground state in the form vibrations and zero-point motion as well as the exciton-phonon scattering in the excited state. For stable, crystalline systems without very light nuclei (hydrogen) the effects of this approximation are minor. Secondly, BSE models X-ray excitation as a single electron-hole pair, which can be problematic for atoms with highly localized d- or f-electrons. 
Thirdly, but most importantly in this work, the conduction electron states are computed using DFT within the local density approximation. The DFT is known to underestimate excitation energies including band gaps, and the excited electron states have an infinite lifetime within the DFT. Higher-level theories, {\it e.g.}, {\it GW} self-energy calculations, can be used to correct the DFT energies, including lifetimes.\\
Lastly, we approximate the many-body photon absorption process as a single-electron interaction and treat the electrons as quasi-particles. These effects can be grouped into an unknown proportionality constant.\\

A comparison of the ocean post-processing results with the reconstructed refractive index from the reflectometry measurements at the O-K edge is shown in Figure~\ref{fig:nk_Oedge}. An atomic thin organic contamination of the quartz surface was assumed for the reflectometry model on which the n\&k fit of the ocean post-processed is based. Omitting the contamination layer in the model results in a significant difference which can best be visualized by comparing the anisotropies.

\begin{figure}[htbp]

\centering
\includegraphics[width=0.4\textwidth]{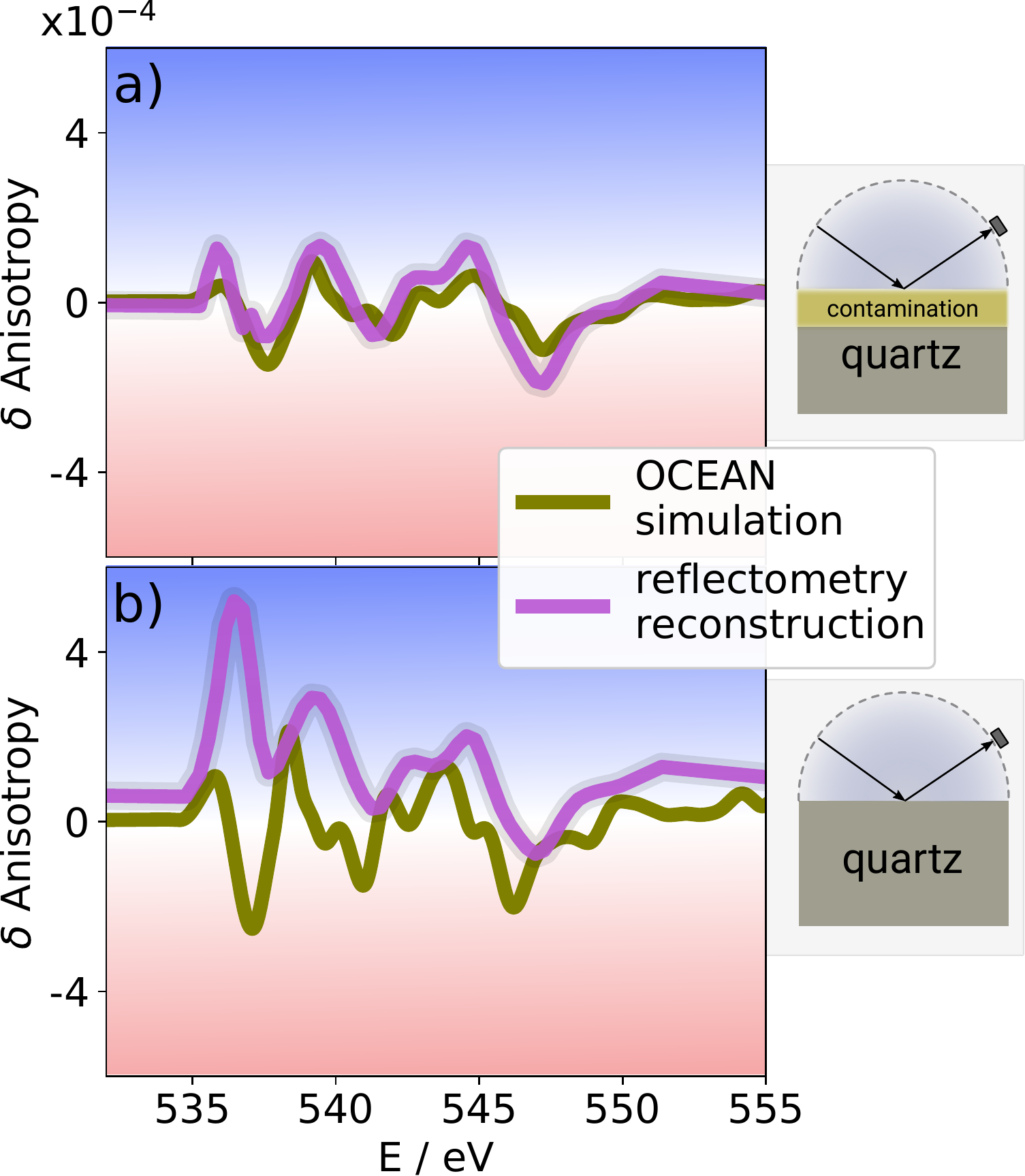}
\caption{The reconstructed anisotropy at the O-K absorption edge of the real part of the refractive index $\delta$ in comparison with a anisotropy obtained from the {\sc ocean} simulation. A reasonable agreement in anisotropy is obtained with a reflectometry model that also allows surface contamination (a).}
\label{fig:nk_aniso}
\end{figure}   

\subsection{\sc{ocean} anisotropy for different reflectometry models}
To allow an unbiased comparison the parameters of the {\sc OCEAN} post-process were adjusted to n\&k values determined by a reflectometry model corresponding to a pure quartz surface. The comparison of the two models is shown in Figure ~\ref{fig:nk_aniso} for the simulated anisotropies reconstructed from the measurement and adjusted on this basis. In the figure the differences for $\delta$ and $\beta$ for both crystal orientations ((100),(001)) are shown as function of the photon energy. The direct comparison of the two models clearly shows that the omission of the contamination layer leads to a significant difference.\\
 
It can be clearly seen that within this simplified model, the reconstruction of the optical constants in the area of the absorption edges is affected. This is also obvious from the fit results of the measured and modeled reflectivities. The anisotropy reconstructed in this way increases significantly and can no longer be correlated with the {\sc ocean} simulation. This observation is on the one hand a proof for the validity of the assumed model, but at the same time it shows how crucial the model error is in the reconstruction of optical constants. First tests based on the Markov Chain Monte-Carlo method showed that the uncertainty for the reconstructed optical constants at the absorption edge increases. Similar observations for the uncertainties of the optical constants are reported by Soufli et al.~\cite{Soufli1997} at the Si-L edge for pure silicon. This probably correlates with the loss in reflectivity at the zero crossing of the real part of the refractive index.

\section{Conclusions}
In this work we have determined optical constants for quartz in a broad photon energy range in the soft X-ray region and can significantly extend existing databases, especially in the vicinity of the Si-L and O-K absorption edges. The comparison with literature values for SiO$_2$ shows a excellent agreement away from absorption edges and the expected deviations at the edges. In these spectral ranges, an anisotropy of the optical constant can be clearly verified depending on the orientation of the crystal. This work could also pave the way to measure possible stress induced anisotropy in thin films by measuring the anisotropy near the absorption edge of the material under consideration (e.g.~Si, O, N, C).\\ The measured anisotropy of quartz is also confirmed by ab initio BSE simulations at the O-K edge using {\sc ocean} calculations. 
These forward calculations also confirm the model assumption of the presence of a thin surface contamination layer for the reconstruction of the optical constants. A perfect agreement between measured and calculated optical constants of quartz around the O-K absorption edge could not yet be achieved with the present theoretical simulations. The BSE approach for the simulation requires many approximations which limit its possibilities. However, a better prediction of fine structures at the edges seems to be possible and offers potential for further chemical analysis if experimental resolution limits are reached.

\section{Acknowledgments}

This project has received funding from the Electronic Component Systems for European Leadership Joint Undertaking under grant agreement No 783247 – TAPES3.
This Joint Undertaking receives support from the European Union's Horizon 2020 research and innovation programme and Netherlands, France, Belgium, Germany, Czech Republic, Austria, Hungary, Israel.
\section{References}
\bibliography{bib}

\end{document}